\documentclass{aa}
\usepackage{lineno}
\usepackage{amsmath}
\usepackage{amssymb}
\usepackage{amsfonts}
\usepackage{comment}
\usepackage{graphicx}
\usepackage{subcaption}
\usepackage{caption}
\usepackage{listings}
\usepackage{hyperref}

\title{Updates on the FRi3D CME model in EUHFORIA}

\author{A. Maharana\href{https://orcid.org/0000-0002-4269-056X}{\includegraphics[scale=0.05]{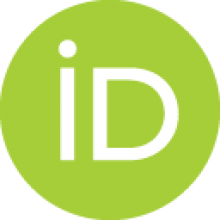}}\inst{1,2} \and K. Plets\inst{1} \and A. Isavnin\inst{3} \and S. Poedts\href{https://orcid.org/0000-0002-1743-0651}{\includegraphics[scale=0.05]{orcid-ID.png}}\inst{1,4} }

 \institute{Centre for mathematical Plasma-Astrophysics, Department of Mathematics, KU Leuven, Celestijnenlaan 200B, 3001 Leuven, Belgium \\
 \and Solar-Terrestrial Centre of Excellence—SIDC, Royal Observatory of Belgium, 1180 Brussels, Belgium\\
 \and Rays of Space, Finland-Belgium \\
  \and Institute of Physics, University of Maria Curie-Sk{\l}odowska, ul.\ Marii Curie-Sk{\l}odowskiej 1, 20-031 Lublin, Poland}
  
\abstract
{
A magnetised flux rope model, "Flux Rope in 3D" (FRi3D) is used in the framework of European Heliospheric Forecasting Information Asset (EUHFORIA) for studying the evolution and propagation of coronal mass ejections (CME). In this paper, we rectify the mistake in the mentioned magnetic field profile of the FRi3D model used in \citet{Maharana2022}, and we clarify the actual profile used in that work. In addition, we provide the recent updates introduced to the FRi3D implementation in EUHFORIA like optimising the "ai.fri3d" python package to reduce computational time and exploring different CME leg disconnection methods to make the numerical implementation more stable.
}

\begin{document}
\maketitle

\section{A non-Lundquist formula}\label{sec:formula_rectification}
It was reported in \citet{Maharana2022} that the FRi3D model implemented in EUHFORIA followed the Lundquist \citep{Lundquist1951} magnetic field profile from \textit{ai.fri3d} package as follows:

\begin{equation}
B_{\rho} = 0 \text{,} \qquad
B_{\varphi} = B_0 J_1(\alpha \rho), \qquad \text{and} \qquad
B_{z} = B_0 J_0(\alpha \rho), \qquad
\end{equation}
where $\rho$ is the poloidal distance from the axis in cylindrical coordinates, $B_0$ is the strength of the core field, $J_0$ and $J_1$ are the Bessel functions of first and second order, respectively, and $\alpha\rho$ gives the first zero of $J_0$ at the edge of the flux rope ($\alpha$ is a free parameter).

However, we noticed that the python package was already updated in 2020 to follow a bi-variate normal distribution, instead of the mentioned Lundquist configuration at the time the work for \citet{Maharana2022} was carried out. In this paper, we clarify the details of the updated FRi3D magnetic field configuration. The bi-variate distribution is able to reproduce the in situ magnetic field profiles in the observed CME events shown in \citet{Maharana2022,Maharana2023,Palmerio2023}. 

\begin{table}[h]
    \centering
    
    \caption{Notation of physical quantities used in the lib.pyx code in \textit{ai.fri3d} package.}
    \begin{tabular}{c|c|c}
    Code variable & Symbol & Physical quantity \\ \hline
    xx[0] & $r$ & radial distance \\ 
    xx[1] & $\theta$ & polar angle \\
    xx[2] & $ \phi$ & toroidal axis angle \\
    xx[3] & $R_t$ & toroidal height \\
    xx[4] & $\varphi_{\text{hw}}$ & half-width \\
    xx[5] & $\varphi_{\text{hh}}$ & half-height \\
    xx[6] & $n$ & flattening \\
    xx[7] & $\theta_p$ & pancaking coefficient \\
    xx[8] & $\tau$ & twist \\
    xx[9] & $\sigma$ & Gaussian standard deviation \\
    xx[10]  & $X$ & $ \int_{-\varphi_{\text{hw}}}^{\varphi_{\text{hw}}} R_t \cos^n(\frac{\pi \phi}{2 \varphi_{\text{hw}}}) d\phi$\\
    \hline
    \end{tabular}
    \label{tab:params}
\end{table}

The magnetic field does not follow Lundquist as reported in \citet{Maharana2022}, but rather is distributed along a bi-variate normal distribution. The distribution is bound by the boundaries of the geometrical shell defined for the CME. The direction of the magnetic field $\mathbf{B}$ is along the magnetic axis line. The magnetic field is primarily defined as:
\begin{equation}
    B(r) = B_0\,\exp\Bigg(-\frac{r^2}{2\sigma^2}\Bigg) = B_0\,\exp\Bigg(-\frac{x^2}{2\sigma^2}-\frac{y^2}{2\sigma^2}\Bigg),
\end{equation}
where $x=r\cos(\theta)$ and $y=r\sin(\theta)$ considering at a particular $\phi$, $B_0$ is the axial magnetic field strength, and $\sigma$ is the standard deviation coefficient of the Gaussian distribution of the total magnetic field in the cross-section of the CME, which is set to a default value of 2. The description of the notation of various symbols is given in Table~\ref{tab:params}. The magnetic field strength ($|B|$) at any point in the flux rope is given by:
\begin{align}
     \frac{|B|}{B_0} &= \exp \left( - \frac{(r\cos{\theta})/r_y)^2}{2 \sigma^2} - \frac{(r\sin{\theta})/r_y)^2}{2 \sigma^2} \right) \times \sin\left(\tan^{-1}  \left( s'(\phi)X \text{ , } \frac{R^2(\phi) 2  \pi \tau }{1 - (1 - \theta_p)  \sin(\theta)} \right)\right) = f_1 \times f_2,
\end{align}
where
$$r_y = R(\phi) \tan( \varphi_\text{hh} ) = R(\phi) R_p /R_t = R_p cos^n{(\frac{\pi}{2} {\phi} / \varphi_{\text{hw}})}  ,$$ $$ s'(\phi) = \sqrt{R(\phi) ^2 + (R'(\phi) )^2}, $$ $$R(\phi) = R_t \cos^n{(\frac{\pi}{2} {\phi} / \varphi_{\text{hw}})} ,$$  and $$X = \int_{-\varphi_{\text{hw}}}^{\varphi_{\text{hw}}} R_t \cos^n(\frac{\pi \phi}{2 \varphi_{\text{hw}}}) d\phi .$$
In the implementation file, $r$ is normalised with $r_y$, which incorporates the toroidal/longitudinal variability in the magnetic field distribution. The above factors can be simplified as follows:
\begin{align}
    f_1 &= \exp \left( - \frac{1}{2}\left(\frac{r}{r_y \sigma}\right)^2 \right), \\
    f_2 &= \sin(\tan^{-1}( A,B   )) = A/\sqrt{A^2 + B^2},
\end{align}
where $A(\phi) = s'(\phi)X$, and $B(\theta,\phi) = \frac{R^2(\phi) 2 \pi \tau}{1 - (1-\theta_p)\sin(\theta)}$. Here, $f_1$ and $f_2$ incorporate the tapering and bending the magnetic axis as per the geometry of the shell. 

\section{Optimisation of calculations}\label{sec:optimisation}
We profiled the \textit{ai.fri3d} package to find the bottlenecks in the code. Then we simplified some computations and discarded some of the repeating computations to optimise the code. \\

\begin{figure}[!htb]
\center
	\includegraphics[width=0.75\linewidth]{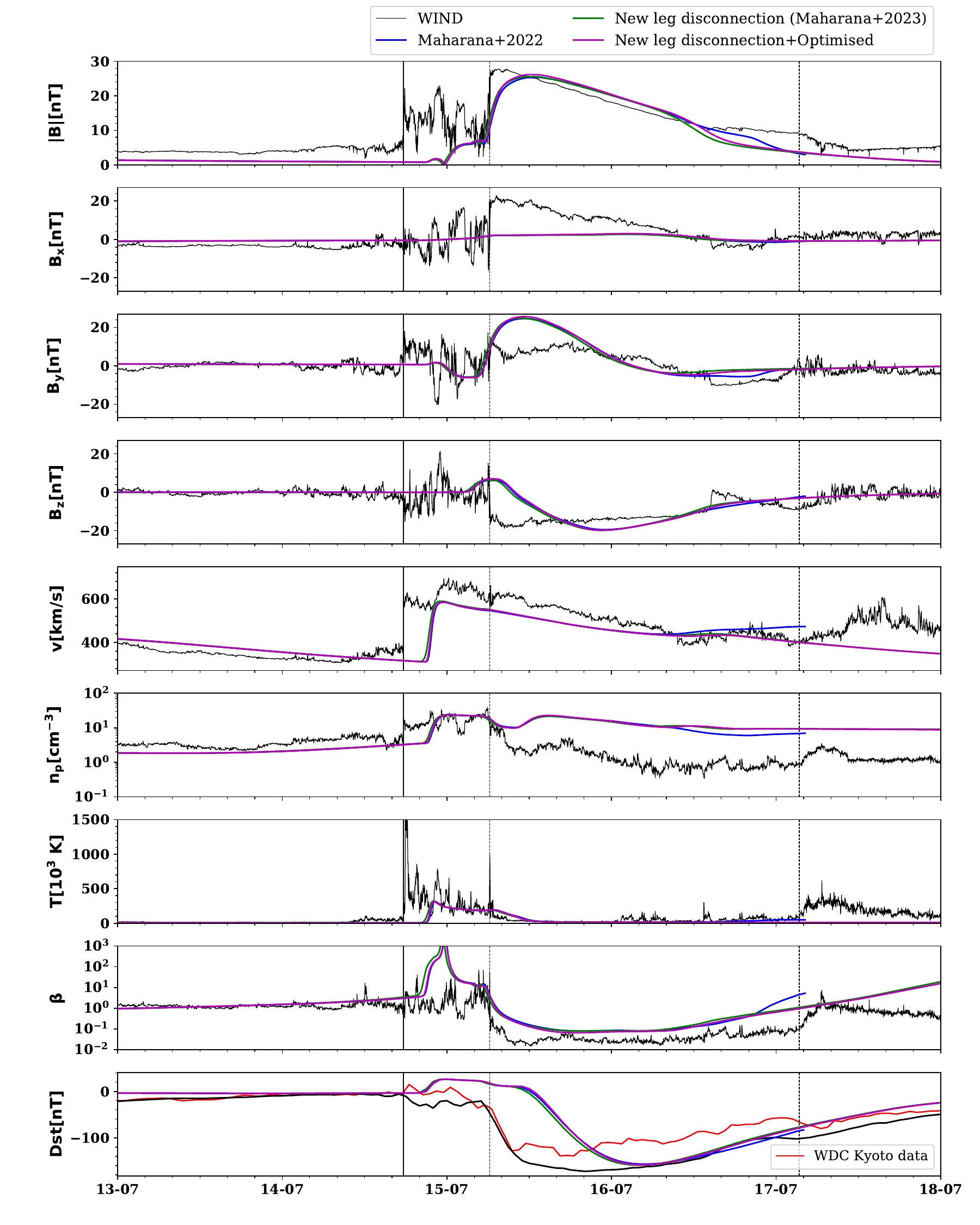}
	\caption{EUHFORIA simulated magnetic field and plasma properties, depicted against the corresponding measurements recorded by the Wind spacecraft. The panels show, from top to bottom: Magnetic field magnitude, magnetic field components in Cartesian coordinates, solar wind bulk speed, proton density, proton temperature, plasma beta, and Dst index \citep[WDC Kyoto data is shown in red and the black profile displays the modeled Dst based on Wind in-situ measurements using the algorithm of][]{obrien2000}. The observed arrival time of the shock is marked with the solid vertical line, and the leading and trailing edges of the ICME ejecta are marked with dashed vertical lines.}
	\label{fig:euh_sims}
\end{figure}

We first simplify the computation of the axial magnetic field strength ($B_0$) which is defined as a function of the longitude $\phi$:
\begin{align}
    B_0(\phi) &= \Phi \left/ \int_{0}^{2\pi} d\theta \int_{0}^{\rho(\theta, \phi)} dr' r' \exp\left(- \frac{1}{2} \left( \frac{r'}{r_y \sigma} \right)^2 \right) \frac{ A }{\sqrt{A^2 + B^2  }} \right.,\\
    &= \Phi \left/ \int_{0}^{2\pi} d\theta \frac{ A }{\sqrt{A^2 + B^2  }}  \int_{0}^{\rho(\theta, \phi)} dr' r' \exp\left(- \frac{1}{2} \left( \frac{r'}{r_y \sigma} \right)^2 \right)  \right. ,
\end{align}
where $\Phi$ is the total magnetic flux constrained from the observations, $$\rho(\theta, \phi) = r_y^2 \alpha_p \left/ \sqrt{r_y^2\cos^2(\theta) + r_y^2\alpha_p^2\sin^2(\theta)}  = r_y \alpha_p \right/ \sqrt{1 -(1 -\alpha_p^2)\sin^2(\theta)} = \alpha_p R_p \frac{\cos^n(\frac{\pi \phi}{2\varphi_\text{hw}})}{\sqrt{1 -(1 -\alpha_p^2) \sin^2(\theta)}}, $$ and $$\alpha_p = 1 - \left[ (1-\theta_p) \left/\sqrt{1 + \left(\frac{n\pi \tan(\frac{\pi \phi}{2\varphi_\text{hw}})}{2\varphi_{\text{hw}}} \right)^2}  \right.\right].$$
Here, $\rho(\theta, \phi)$ is the radial coordinate in the local FRi3D coordinates, and $\alpha_p$ is a longitudinal implementation of pancaking ($\theta_p$) in the flux rope. Finally, $B_0$ is multiplied with the unit position vector of the magnetic axis at each point to obtain the magnetic field vector. First, we reduced the 2D integral in the computation of $B_0$ into a 1D integral by analytically integrating the Gaussian function:
\begin{align}
    B_0(\phi)&= \Phi \left/ \int_{0}^{2\pi} d\theta \frac{(r_y \sigma)^2 A \left[1-\exp\left(- \frac{1}{2} \left( \frac{\rho(\theta,\phi)}{r_y \sigma} \right)^2 \right) \right] }{\sqrt{ A^2 + B^2 }}  \right.,\\
    &= \frac{\Phi}{ s'(\phi)X (1 - (1 - \theta_p) (r_y \sigma)^2} \left/  \int_{0}^{2\pi} d\theta \sin(\theta)  \frac{1-\exp\left(- \frac{1}{2} \left( \frac{\rho(\theta,\phi)}{r_y \sigma} \right)^2 \right)}{\sqrt{ \left(R^2(\phi) 2  \pi \tau \right)^2 + \left[ s'(\phi)X  (1 - (1 - \theta_p)  \sin(\theta)) \right]^2  }}   \right..
\end{align} 
This step contributed significantly in reducing the computation time in EUHFORIA simulations as the double integral had to be computed at at every point in the CME cross-section through the inner boundary throughout its injection.\\

The second optimisation was achieved by defining a separate function `integralphi' to compute the coordinates of the magnetic field axis. This method replaced the repeated computation of the integral in $X$ (mentioned in Section~\ref{sec:formula_rectification}) in the functions to compute the twist or axial magnetic field required for the computation of magnetic field components. Instead, the result of the integral was efficiently used in the above-mentioned functions. This step contributed significantly in reducing the computation time in EUHFORIA simulations as the integral had to be computed at multiple times while checking the position of grid points on the EUHFORIA inner boundary with respect to the FRi3D magnetic axis. These optimisations are implemented in the latest version of \textit{ai.fri3d} package version 0.0.17 (\url{https://pypi.org/project/ai.fri3d/}). Using the updated package to obtain the magnetic field configuration at a particular point takes almost 60$\%$ less time. The EUHFORIA simulation with the optimised version takes $\sim 55\%$ less time for the case shown in \citet{Maharana2022}, and we obtain the nearly the same CME arrival time and magnetic field profiles at Earth, as shown in Fig.\ref{fig:euh_sims} and discussed below.

\section{Updates on the FRi3D implementation in EUHFORIA}
\subsection{CME leg disconnection}
The FRi3D CME leg disconnection follows the process in which the solar wind plasma and magnetic field variables are reassigned at the inner boundary of the EUHFORIA heliosphere within the cross-section where the CME was injected. In \citet{Maharana2022}, the CME legs were disconnected point by point when the CME speed dropped below the ambient solar wind speed in the grid points where the CME was injected. We introduced a new leg disconnection method whose numerical implementation consumes less computational time and is more stable. The flux rope was fully disconnected from the boundary (i.e., all points in the flux rope cross-section were replaced with solar wind properties at the same time) when the magnetic axis of the CME reached 1~au. 
The updated method has been used in \citet{Maharana2023,Palmerio2023} to obtain satisfactory results.

\subsection{Addition of skew parameter in EUHFORIA implementation}
We included the skew parameter of the FRi3D model in the framework of EUHFORIA. The implementation was tested to model the propagation of a flux rope resulting from an extended asymmetric filament eruption as presented in \citet{Lynch2021}. The results of EUHFORIA simulation with FRi3D model are reported in \citet{Palmerio2023}. The FRi3D model performed better than the CME models with spherical/spheroidal geometry in predicting CME arrival time and magnetic field upon impact.

\section{Discussion}
The results of EUHFORIA simulations with different versions of FRi3D implementation are shown in Fig.\ref{fig:euh_sims}. All the simulations were performed on 144 parallel processors of the Vlaams Supercomputer Centrum (VSC), a Belgian high-performance computing facility\footnote{\url{http://www.vscentrum.be}}. The time series shown by the legend {Maharana+2022} (blue profile) is the first ever implementation of FRi3D in EUHFORIA, and consumed a wall-clock time of 24~hours to produce a forecast of 5 days. Updating the leg disconnection method, the simulation was completed in 9 hours to produce a forecast of 7 days (green profile). Adding the optimisations discussed in Section~\ref{sec:optimisation}, the forecast of 7 days was achieved in only 4 hours and 9 minutes, that is $\sim55\%$ less computational time expenditure (magenta profile). The two updated versions of FRi3D implementation are qualitatively similar to the first implementation in \citet{Maharana2022}, and are much more stable and less time consuming.

\bibliographystyle{aa}

\end{document}